\documentclass[12pt,twoside]{article}

\usepackage{rotating} 

\usepackage{setspace}
\usepackage{fancyhdr}
\usepackage[margin=1in, headheight=15pt]{geometry}
\usepackage{lipsum}
\usepackage{epigraph}
\usepackage{titlesec}
\titlespacing*{\subsubsection}{0pt}{6pt plus 7pt minus 2pt}{2pt plus 2pt minus 2pt}

\usepackage[dvipsnames,table]{xcolor}

\usepackage{graphics}
\usepackage{tikz}
\usetikzlibrary{calc}
\usetikzlibrary{decorations.pathreplacing}
\usetikzlibrary{arrows.meta}
\tikzset{every picture/.style={/utils/exec={\sffamily}}}
\usepackage{amsmath,amsfonts,amsthm}
\usepackage{mathrsfs}
\usepackage{cancel}
\usepackage[normalem]{ulem}
\usepackage{siunitx}
\usepackage{float}
\usepackage{enumitem}
\usepackage{colortbl}
\usepackage{subcaption}
\usepackage[font=footnotesize,labelfont=bf]{caption}
\usepackage{tabularx}
\newcolumntype{L}[2]{>{\hsize=#1\hsize\columncolor{#2}\raggedright\arraybackslash}X}%
\newcolumntype{R}[2]{>{\hsize=#1\hsize\columncolor{#2}\raggedleft\arraybackslash}X}%
\newcolumntype{C}[2]{>{\hsize=#1\hsize\columncolor{#2}\centering\arraybackslash}X}%
\usepackage{wrapfig}
\usepackage{transparent}
\usepackage{adjustbox}
\usepackage{svg}
\usepackage{multirow}
\usepackage[hang,flushmargin]{footmisc}

\usepackage{color}
\definecolor{mygreen}{rgb}{0,0.6,0}
\definecolor{mygray}{rgb}{0.5,0.5,0.5}
\definecolor{mymauve}{rgb}{0.58,0,0.82}
\definecolor{mylightgray}{gray}{.9}
\definecolor{myblue}{rgb}{.28,.24,.55}
\definecolor{mylightblue}{rgb}{0.74, 0.83, 0.9}
\definecolor{mybrightred}{rgb}{1,.13,.32}
\definecolor{mypink}{rgb}{0.96, 0.76, 0.76}
\definecolor{mylightgreen}{rgb}{0.66, 0.89, 0.63}
\usepackage{colorspace}

\definecolor{ddarkgreen}{HTML}{004166}
\definecolor{dmediumgreen}{HTML}{4dbeff}
\definecolor{dlightgreen}{HTML}{ccecff}
\definecolor{ddarkpurple}{HTML}{621a04}
\definecolor{dmediumpurple}{HTML}{fb7850}
\definecolor{dlightpurple}{HTML}{fdd9ce}
\definecolor{dlightgrey}{HTML}{DCDCDC}
\definecolor{o1}{HTML}{40004b}
\definecolor{o2}{HTML}{7f3c8d}
\definecolor{o3}{HTML}{ad8abd}
\definecolor{o4}{HTML}{dec8e2}
\definecolor{o5}{HTML}{f7f7f7}
\definecolor{o6}{HTML}{ccebc6}
\definecolor{o7}{HTML}{80c481}
\definecolor{o8}{HTML}{2b8642}
\definecolor{o9}{HTML}{00441b}
\definecolor{llightpurple}{HTML}{f8e8fc}
\definecolor{llightyellow}{HTML}{fdf6e7}

\usepackage{sectsty}
\sectionfont{\normalfont\fontsize{13.5}{15}\bfseries}
\subsectionfont{\normalfont\fontsize{13.5}{15}\bfseries}
\subsubsectionfont{\normalfont\fontsize{13.5}{15}\bfseries}

\usepackage{natbib}
\setcitestyle{aysep={}}

\usepackage{hyperref}

\hypersetup{colorlinks=true, citecolor=gray, linkcolor=gray, urlcolor=gray}

\usepackage{lmodern}
\usepackage{listings}
\usepackage[textsize=tiny, backgroundcolor=mypink, linecolor=red]{todonotes}

\makeatletter
\newcommand{\myfnsymbol}[1]{%
  \expandafter\@myfnsymbol\csname c@#1\endcsname
}
\newcommand{\@myfnsymbol}[1]{%
  \ifcase #1
  \or 1
  \or 2
  \or 3
  \or 4
  \or 5
  \or 6
  \or 7
  \or 8
  \or 9
  \or 10
  \or 11
  \or 12
  \or \TextOrMath{\textasteriskcentered}{*}
  \or \TextOrMath{\textdagger}{\dagger}
  \or \TextOrMath{\textdaggerdbl}{\ddagger}
  \fi
}
\newcommand{\affiliationA}{\@myfnsymbol{1}}
\newcommand{\affiliationB}{\@myfnsymbol{2}}
\newcommand{\affiliationC}{\@myfnsymbol{3}}
\newcommand{\affiliationD}{\@myfnsymbol{4}}
\newcommand{\affiliationE}{\@myfnsymbol{5}}
\newcommand{\affiliationF}{\@myfnsymbol{6}}
\newcommand{\affiliationG}{\@myfnsymbol{7}}
\newcommand{\affiliationH}{\@myfnsymbol{8}}
\newcommand{\affiliationI}{\@myfnsymbol{9}}
\newcommand{\affiliationJ}{\@myfnsymbol{10}}
\newcommand{\affiliationK}{\@myfnsymbol{11}}
\newcommand{\affiliationL}{\@myfnsymbol{12}}
\newcommand{\titlemark}{\@myfnsymbol{13}}
\newcommand{\equalcontrib}{\@myfnsymbol{14}}
\newcommand{\corresp}{\@myfnsymbol{15}}
\makeatother

\newcommand{\titlenote}{This work is funded by Academy of Finland grants 320780 and 320781. We thank Aasa Karimo, Otto Schultz, and Amirmohammad Ziaei Bideh for providing invaluable research assistance. We acknowledge the computational resources provided by the Aalto Science-IT project. Our data collection protocol and its code implementation can be found at \href{https://github.com/tedhchen/componMultilayer}{\texttt{github.com/tedhchen/componMultilayer}}.}
\newcommand{\mytitle}{\vspace{-2cm}Climate Policy Elites' Twitter Interactions across Nine Countries\textsuperscript{\titlemark}}
\newcommand{\shorttitle}{Climate Policy Elites' Twitter Interactions across Nine Countries}
\author{
Ted Hsuan Yun Chen\textsuperscript{\affiliationA,\equalcontrib,\corresp} \and
Arttu Malkamäki\textsuperscript{\affiliationB,\equalcontrib,\corresp} \and
Ali Faqeeh\textsuperscript{\affiliationB} \and
Esa Palosaari\textsuperscript{\affiliationB} \and
Anniina Kotkaniemi\textsuperscript{\affiliationC} \and
Hasti Narimanzadeh\textsuperscript{\affiliationB} \and
Laura Funke\textsuperscript{\affiliationD} \and
Cáit Gleeson\textsuperscript{\affiliationE} \and
James Goodman\textsuperscript{\affiliationF} \and
Antti Gronow\textsuperscript{\affiliationC} \and
Marlene Kammerer\textsuperscript{\affiliationG} \and
Myanna Lahsen\textsuperscript{\affiliationH} \and
Alexandre Marques\textsuperscript{\affiliationH} \and
Petr Ocelik\textsuperscript{\affiliationI} \and
Shivangi Seth\textsuperscript{\affiliationJ} \and
Mark Stoddart\textsuperscript{\affiliationD} \and
Martin Svozil\textsuperscript{\affiliationI} \and
Pradip Swarnakar\textsuperscript{\affiliationJ} \and
Matthew Trull\textsuperscript{\affiliationF} \and
Paul Wagner\textsuperscript{\affiliationK} \and
Yixi Yang\textsuperscript{\affiliationD} \and
Mikko Kivelä\textsuperscript{\affiliationB} \and
Tuomas Ylä-Anttila\textsuperscript{\affiliationC,\corresp}
}

\newcommand{\surname}{Chen \& Malkam\"aki et al.}

\title{\Large\mytitle}
\date{June 9, 2026\vspace{-1cm}}

\begin{document}
\pagenumbering{roman}
\singlespacing

\renewcommand{\thefootnote}{\myfnsymbol{footnote}}
\maketitle\thispagestyle{empty}
\footnotetext[13]{\titlenote}%
\footnotetext[1]{Department of Environmental Science and Policy, George Mason University, USA}%
\footnotetext[2]{Department of Computer Science, Aalto University, Finland}%
\footnotetext[3]{Faculty of Social Sciences, University of Helsinki, Finland}%
\footnotetext[4]{Memorial University of Newfoundland, Canada}%
\footnotetext[5]{Dublin City University, Ireland}%
\footnotetext[6]{University of Technology Sydney, Australia}%
\footnotetext[7]{University of Bern, Switzerland}%
\footnotetext[8]{The National Institute for Space Research (INPE), Brazil}%
\footnotetext[9]{Masaryk University, Czech Republic}%
\footnotetext[10]{Indian Institute of Technology Kanpur, India}%
\footnotetext[11]{Edinburgh Napier University, United Kingdom}%
\footnotetext[14]{Equal contribution alphabetically ordered by surname for the version of record. Both authors are free to list themselves as first author in personal publication lists.}%
\footnotetext[15]{Corresponding. ted.hsuanyun.chen@gmail.com, arttu.1.malkamaki@aalto.fi, tuomas.yla-anttila@helsinki.fi.}%

\maketitle
\thispagestyle{empty}
\begin{abstract}
    \noindent
    Social media is an important space for interactions between climate policy actors, with a burgeoning literature recognizing them as critical platforms of political contestation. We identified Twitter accounts associated with 904 climate change policy actors across nine countries, and collected their activities from 2017–2022, totalling 40 million activities from 16,086 accounts at different organizational levels. We studied these actors and their interactions as a polycentric governance system, emphasizing how boundary blurring between the public and private on social media platforms uniquely shapes the online policy process. Initial results show there is considerable temporal and cross-national variation in how prominent climate-related activities were, but all national policy systems generally responded to climate-related events, such as climate protests, in a similar manner. Examining patterns of interaction within and across countries, we find that these national policy systems rarely directly interact with one another, but are connected through consistently engaging with the same content produced by accounts of international organizations, climate activists, and researchers.

    \vspace{0.5cm}
\end{abstract}

\clearpage

\thispagestyle{fancy}
\tableofcontents

\newpage
\pagenumbering{arabic}
\onehalfspacing

\section{Introduction}
Social media is an important space for interactions between climate policy actors \citep{dellmuth2023climate,kavada2022environmental}, with a burgeoning literature recognizing them as critical platforms of political contestation and democratic participation \citep{robertson2019democratic,theocharis2015conceptualization}. Many organizations and individuals key to climate policy actively participate on social media platforms (e.g., UNFCCC, Greta Thunberg), using them to engage with the public and with one another. Twitter (now X), in particular, is among the most important social media platforms (at least until Elon Musk's takeover in 2022) where policy actors and other political elites engage in various aspects of the policy process \citep{conway2015rise,ausserhofer2013national}, such as setting agendas and discourse \citep{gilardi2022social}, vying for political support \citep{mcgregor2017twitter}, and mobilizing movements \citep{segerberg2011social}.

There is an extensive literature on climate politics and climate policy debates on Twitter \citep{fownes2018twitter}, but researchers have yet to explicitly consider Twitter or other social media platforms through the lenses of the policy and governance literatures \citep{ostrom2010polycentric,atkinson1992policy}, specifically, as sociotechnical policy systems where policy elites collaborate and compete to influence the climate policy process while operating under the platforms' affordances and constraints \citep{dellmuth2023climate}.

First, studies usually do not focus on the policy \textit{system}, often missing important components of the policy contestation process \citep{cairney2012complexity}. They either do not disaggregate policy actors from the general public, looking instead at general trends in climate communication or opinion dynamics \citep{falkenberg2022growing,chen2021polarization,cody2015climate}, or only examine an incomplete subset of climate policy actors, focusing on specific types\footnote{Such as politicians \citep{ebrey2020twitter,ghoraba2023influential}, non-governmental organizations \citep{hart2024climate,vu2020leads}, and scientists \citep{walter2019scientific}.} or using system bounding rules that exclude actors based on how active or popular they are \citep{goritz2022international,stier2018activists}.

Second, studies rarely consider the policy system's \textit{sociotechnical} qualities despite the importance of contextual effects to how policy actors interact \citep{metz2023policy}. From the technology affordances literature \citep{evans2017explicating}, we know that social media platforms have distinct affordances -- perceived or understood properties of the technology that enable and constrain action -- that shape the nature of social interactions \citep{ronzhyn2023defining}.
With recent studies showing that online and offline policy spaces follow different organizing principles, manifesting in observable differences in policy process interactions and outcomes \citep{kotkaniemi2024policy,malkamaki2023complex,hayes2018multiplex},
can we examine the affordances that are activated in the Twitter climate policy system to see how they shape the policy process?

Among different social media affordances \citep{ronzhyn2023defining}, we focus on the potential for blurred boundaries between personal and professional lives on social media \citep{siegert2019online}, which introduces both substantive and methodological implications for how we understand this policy system. Policy actors are often organizational, comprising various subunits and employees or members, all likely holding different preferences and understandings of their organizations' goals and coordination modes \citep{meglino1998individual,van1976determinants}. Because social media platforms are populated by accounts with clearly-indicated professional affiliations that are at the same time for personal use \citep{ford2011reconceptualizing}, the existence of within-organization variation raises the critically under-explored question about how we conceptualize policy \textit{actors} and policy contestation on social media.

Despite this, studies of policy actors on social media overwhelmingly focus on only the official account of organizations, leaving the entirety of policy actions undertaken by organizations' executives and rank-and-file members unexamined.
With this in mind, we focus our examination on the disaggregation of policy actors on social media and how systematic variation across organizational levels plays into the organizing principle of this policy space.
For example, while these non-official accounts heighten the visibility of the organizations they belong to, they also behave differently from these organizations' official accounts, including how they align with other policy actors on social media, which calls into question organizational policy actors' messaging coherence \citep{palosaari2024my}.

In highlighting these platformed affordances, we directly engage with the policy and governance literatures by explicitly bringing the concept of polycentric governance to social media policy research. Seminal studies have clearly demonstrated that governance systems are best considered as polycentric, comprising policy actors operating across different levels and forming different loci of authority \citep{ostrom2010polycentric}. By examining Twitter accounts from different levels of climate policy organizations, we springboard from this notion and show that the complex nature of multiscalar policy behavior does not stop at the boundaries of organizational entities. Here, we explore the tension between the flatness of Twitter as a social media platform and the oft-observed existence of social hierarchies in interactional and communicative behaviors \citep{kotkaniemi2024policy,dagoula2019mapping}.
Again, this links back to the importance of understanding social media affordances and their relationship to policy organizations \citep{behrend2024implications}, as the highly visible and persistent nature of social media platforms \citep{neubaum2022s,ramirez2018social}, which allows for contemporaneous and archival monitoring, potentially makes hierarchical organizational constraints more salient despite the technically flat platform.

To conduct our exploration, we identified the relevant Twitter accounts for 904 climate policy actors across nine countries -- Australia, Brazil, Canada, Czechia, Finland, Germany, India, Ireland, Sweden. These policy actors -- including government ministries, NGOs, businesses, and research institutes -- were identified by local experts, and represented the most important organizational entities in each country's climate policy system \citep{yla2018climate}. For each policy organization, we identified 1) their primary Twitter account, 2) the accounts of their top-level executives, 3) the accounts associated with their organizational subunits or specific functions, and 4) the accounts of their rank-and-file members. Then we collected the Twitter activities of all these accounts from 2017--2022, totalling to 40 million activities across 16,086 accounts. In the first iteration of our paper, we conducted exploratory analysis on how these accounts interacted with one another within and across the boundaries of their national policy systems. We discuss planned analysis at the end of our manuscript.

\section{Results}
\begin{figure}
    \centering
    \includegraphics[width=1\linewidth]{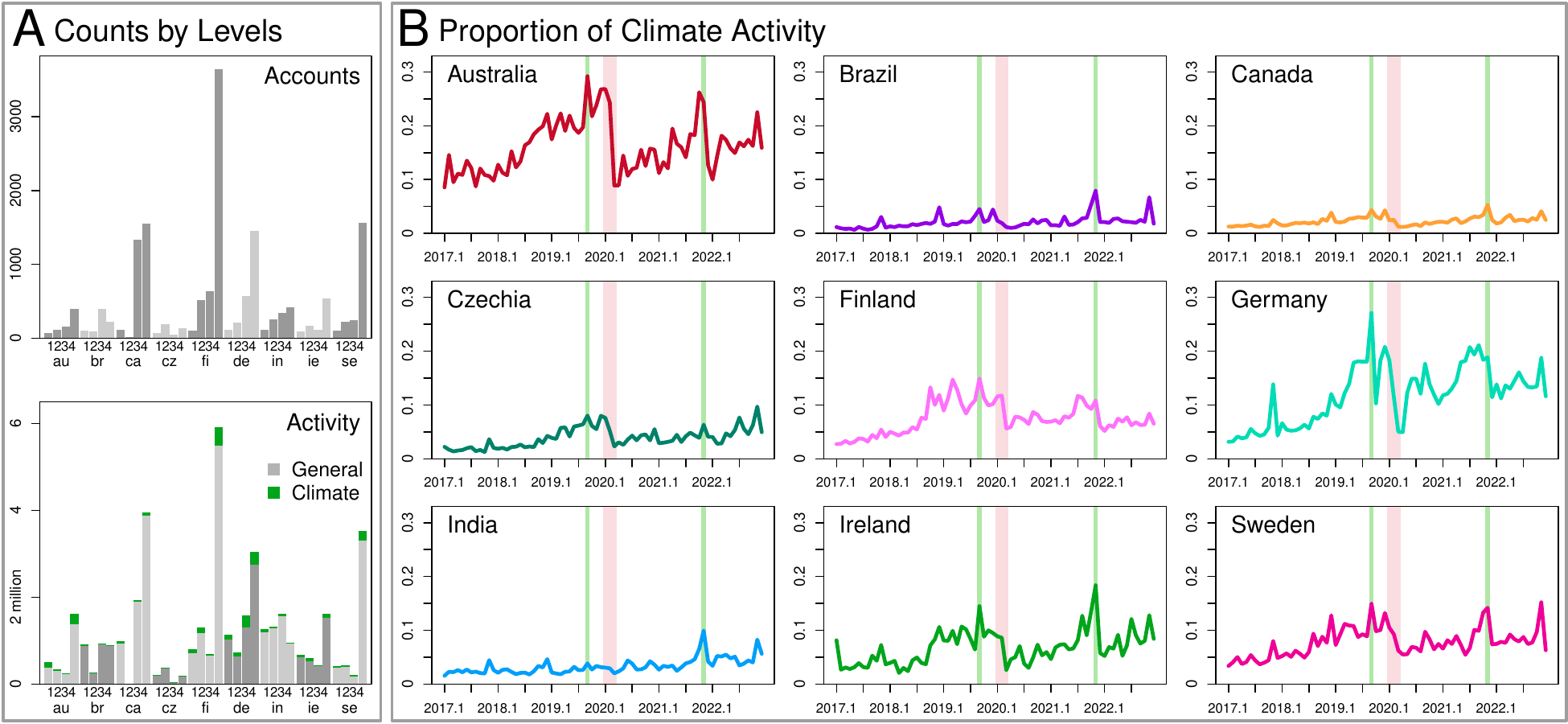}%
    \caption{Overview of our data. Panel A) Numbers of accounts and activity by country and organization level. Panel B) Monthly ratio of climate activity to total activity. The first green bar marks the Fridays for Future strikes in September 2019, and the second green bar marks COP26 in November 2021. The red bar covers January--March 2020, which is the onset of the COVID-19 pandemic.}
    \label{fig:overview}
\end{figure}

\subsection{System Size and Activity Levels}
As shown in \autoref{fig:overview} (Panel A), there is considerable variation in the number of identified accounts and their total activity across countries and organization levels. Naturally, based on our definition of the four different organization levels, we identified many more organizational subunits and rank-and-file members or employees than we did organization's executives. Interestingly, this distribution did not fully translate to activity counts. While rank-and-file members are collectively the most active simply due to their numbers, organizations' primary accounts tend to be just as active, if not more, when compared to the activity levels of executive and organizational subunits.

Of our total 40 million activities, 2.7 million (or 6.8\%) are climate change-related based on a relatively strict dictionary-based classification (see \autoref{sec:dictionary}). As shown in \autoref{fig:overview}~(Panel B), the proportion of climate tweets to general tweets, which captures the relative importance and salience of the climate change issue, varied greatly both by country and across time. For example, in September 2019 (marked with the first green bar), which was the time of the Fridays for Future climate strikes, the Australian policy system had 28\% of climate-related activity, the highest observed proportion in our data set. The cross-national variation suggests that climate policy organizations had different levels of specialization across different national policy systems, and the temporal variation means that climate policy communication on Twitter responds to broader societal factors.

Indeed, while the ratio of climate activity to overall activity varied across each national system, the temporal variation in general and climate Twitter activity exhibits a number of interesting similarities and is generally comparable across systems. This indicates that they are responsive to a similar set of events. For example, we see climate activity peaks during the 2019 Fridays for Future climate strikes and during COP26 in November 2021. Most notably, climate activity was subject to a COVID dip, whereby the world's focus on COVID-19 shifted attention away from climate change \citep{smirnov2022covid}. 

\subsection{Interactions within and across countries}
To understand the extent to which climate policy actors engage with one another across policy systems, we examined different types of interactions between accounts. First, we looked at direct connections, which we measured using direct retweeting. Second, we looked at indirect connections, which we measured as joint retweets of a third-party tweet.

\begin{figure}[!h]
    \centering
    \includegraphics[width=1\linewidth]{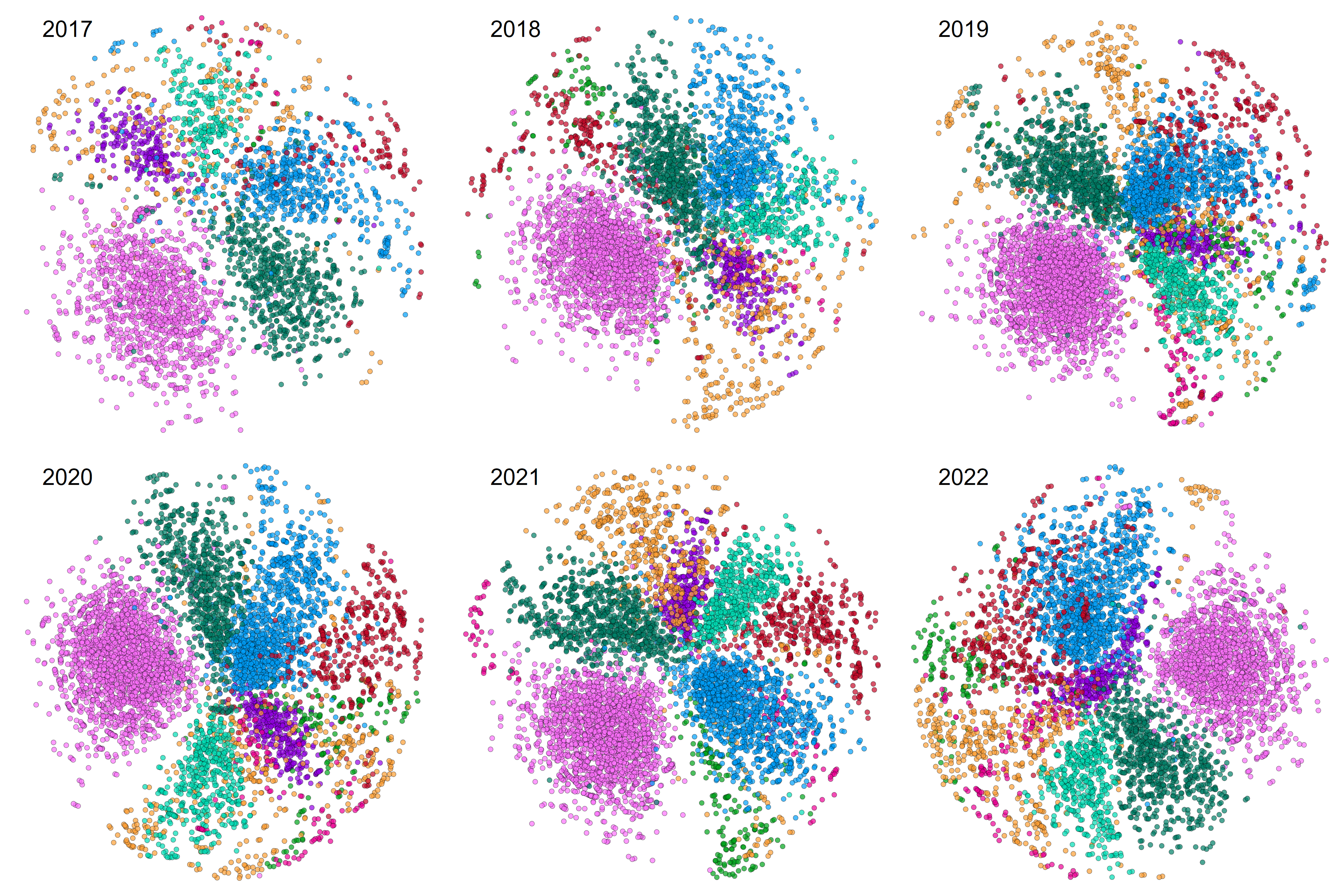}
    \includegraphics[width=1\linewidth]{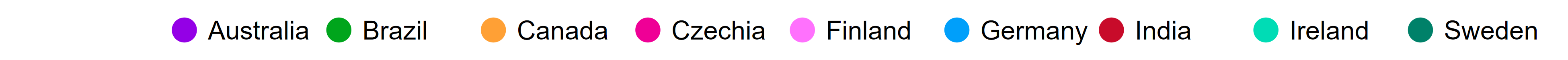}
    \caption{Climate change retweet networks, 2017--2022. Nodes are placed using the Fruchterman-Reingold algorithm and edges are not plotted for visual clarity. Only the largest weakly connected component is shown.}
    \label{fig:networks}
\end{figure}

\subsubsection{Direct Engagement}
\begin{figure}[!h]
    \centering
    \includegraphics[width=0.5\linewidth, page = 1]{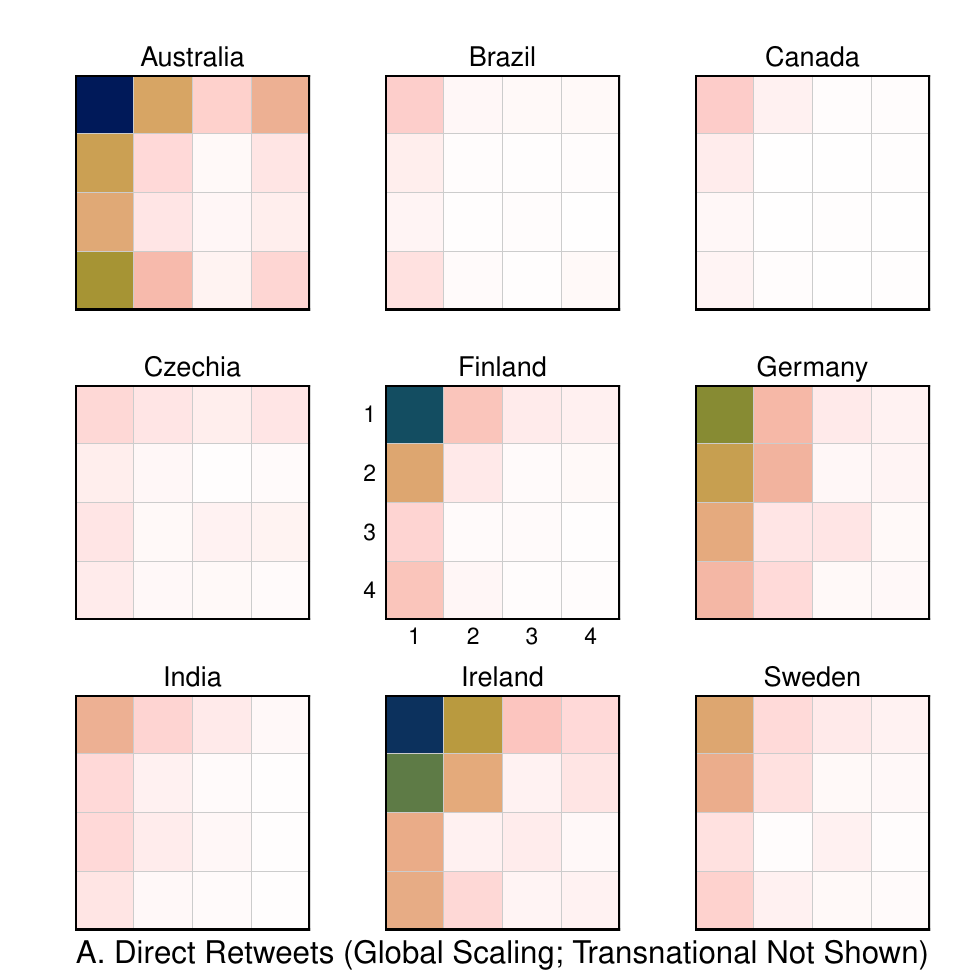}%
    \includegraphics[width=0.5\linewidth, page = 2]{mixing_combined.pdf}
    \includegraphics[width=0.5\linewidth, page = 3]{mixing_combined.pdf}%
    \includegraphics[width=0.5\linewidth, page = 4]{mixing_combined.pdf}

    \caption{Mixing matrices for direct (retweets) and indirect (joint retweeting) connections. Cells are the mean logged ($e$) counts of connections (with min-max scaling) between all pairs of accounts with the given country-level combinations. Darker cells indicate higher connectedness. Axes labels refer to 1) organization primary, 2) individual executive, 3) organization auxiliary, and 4) individual non-executive accounts.}
    \label{fig:mixing}
\end{figure}

\autoref{fig:networks} shows the annual climate change retweet network. It is immediately apparent how isolated each national policy system is from one another, with some minor exceptions. \autoref{fig:mixing} shows different types of mixing matrices pooled across time. Panel A clearly reflects the network plots, with the Australian and Irish systems most densely connected, and all national systems much more densely connected internally than across systems. In Panel B, we rescale the relative density to within each national policy system, which shows a number of similarities across countries. First, policy actors' primary organization accounts tend to be more connected to one another, and also tend to be both more active in retweeting and more popularly retweeted. Next, in Panel C, we examine the cross-country connections by themselves. We find that Germany and Sweden tend to be active in retweeting across the boundaries of their national policy systems, with Swedish executive accounts being the most active. These cross-country patterns are also clearly driven by regional ties, such as the considerably higher amount of reciprocity between Swedish and Finnish accounts.

\subsubsection{Indirect Engagement}
Next, we consider indirect connections between policy actors. We measure this by considering instances of accounts jointly retweeting the same tweet made by an account not on our policy actor roster. Regardless of whether these joint retweets are the result of direct information diffusion, they indicate similar preferences among connected accounts (i.e., because they endorsed the same content) and similar resulting information diffusion (i.e., the same external information was shared into the actors' networks).
Here, relative to direct retweets, Australia is the most connected, followed by Germany and Ireland. Geographical and cultural factors ostensibly plays a role here, as India and Brazil have the least connected systems.

\begin{figure}
    \centering
    \includegraphics[width=\linewidth]{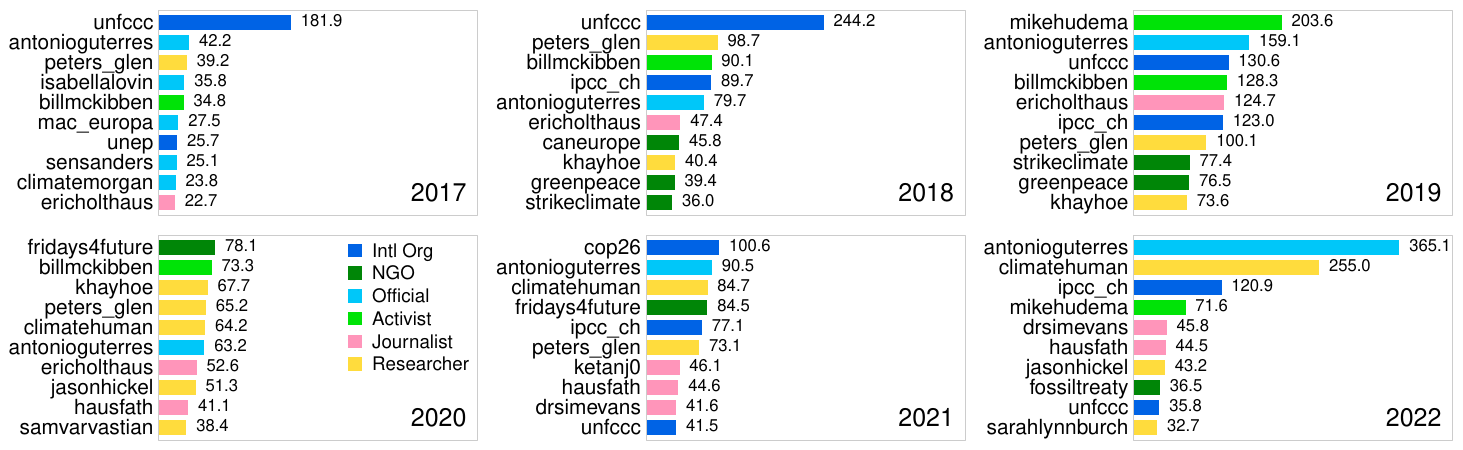}
    \caption{Top ten bridging accounts by year. The bridging statistic is the sum of the logged ($e$) cross-country joint retweet counts across all retweeted tweets from the account in the given year.}
    \label{fig:bridging}
\end{figure}

To better understand these indirect connections across countries, we examine the bridging actors, or accounts whose tweets are most jointly retweeted by accounts from different countries. For each tweet, we calculated its bridging statistic, which is the number of times the tweet was retweeted by pairs of accounts from different countries. To prioritize accounts that consistently bridge at a high volume, as opposed to those that had a single tweet become massively viral, we took the natural log of all tweets' bridging statistics then summed them to the account level.\footnote{
$\sum_{t \in T}\ln(\text{bridging}_t)$%
, where $t$ is a single tweet out of all the account's tweets $T$, and the bridging statistic is the count of cross-country account pairs jointly retweeting that tweet.
}
As shown in \autoref{fig:bridging}, these bridging actors tend to be the accounts of international organizations, including senior officials from these agencies, and individual climate activists. Interestingly, during 2020, scientists were overwhelmingly represented in the list of top bridging actors.

\section{Future Work}
In the current study, we have shown a set of descriptive statistics about the nine national climate policy systems that is part of a larger transnational climate policy system on Twitter. In future iterations of this study, we will expand our analysis both in terms of conducting deeper analysis into policy actors' behaviors and interactions, and in terms of considering a broader set of observable behaviors.

First, while we have shown general interaction patterns, we have not statistically examined the generative features of these relational systems. We will statistically model the structure of their interaction networks based on which national policy system they belong to, their sectors, and the specific organizational level of each account. Second, while we have primarily looked at retweeting patterns, they only constitute a portion of interactions between policy actors on Twitter. We will extend our analysis to include other interactions such as direct responses to each other or directly linking to each other's tweets. Third, in terms of analyzing the bridging actors, showing the strongest bridging actors is only part of the picture, as it misses other individuals and organizations ostensibly important to climate policy Twitter but do not bridge national policy systems. Finally, we will also examine the textual content of the tweets to understand whether the literal text is used semantically differently across different contexts and by different users \citep{rodriguez2023embedding}.

\section{Methods and Data}
\subsection{Identifying Climate Policy Elites in National Policy Systems}
We collected the roster of relevant policy elites of each country's climate policy system as part of the Comparing Climate Change Policy Networks project \citep{yla2018climate}. For each country, a team of local researchers identified 50--100 of the most influential organizations in the field of climate change policy based on the team's local knowledge of the case, media appearances of the organizations, and consultations with local climate policy experts. There are a total of 1045 organizations across the nine countries. These organizations are policy actors with an interest in influencing climate policy, and have employees or members who specialize in climate policy issues. These organizations represent different sectors of societal actors, including governmental actors, political parties, NGOs, businesses, and science actors, and can be considered as representing the climate policy elite in each of the countries.

\subsection{Identifying Policy Organizations and Members on Twitter}
For each policy actor, we identified four ``levels'' of organizational roles: 1) organization primary, 2) individual executive, 3) organizational auxiliary, and 4) individual non-executive. These levels are summarized in \autoref{tab:levels}, using WWF-Australia as an example.
Here, we describe the conceptual bounds of each organizational level, and describe how we identified each of these subsets of accounts. A schema of our protocol is shown in \autoref{fig:protocol}.\footnote{Additional details of our account identification and data collection protocols can be found in our codebook at the following Github repository (\href{https://github.com/tedhchen/componMultilayer}{\texttt{github.com/tedhchen/componMultilayer}}).} Of the 1045 organizations, we were able to identify at least one Twitter account for 904 (87\%).

\begin{table}[b]
    \centering\renewcommand{\arraystretch}{1.5}\footnotesize{
    \begin{tabularx}{\textwidth}{>{\hsize=.143\hsize}X >{\hsize=.473 \hsize}X >{\hsize=.384\hsize}X}
    \hline \hline
    Level & Description & Example \\ \hline
    Organization Primary & Primary organizational account of the policy actor. & \texttt{@wwf\_australia} \\
    Individual Executive & Accounts of the organization's executive personnel. This level includes only the top-level executive and the chair of the board of directors. & \texttt{@rachlowry} \linebreak(Chief Conservation Officer) \\
    Organization Auxiliary & Accounts of organization's subunits or accounts used for focused functions of the organization. & \texttt{@rewildingoz} \linebreak (Rewilding Program)\\
    Individual Non-executive & Accounts of the organization's personnel who are not members of the top-level executive. & \texttt{@darrengroverwwf} \linebreak (Head of Healthy Land and Seascapes) \\ \hline
    \end{tabularx}
    \caption{Overview of different organizational levels. Examples are from WWF-Australia.}\label{tab:levels}}
\end{table}

\begin{figure}[!htb] 
    \centering 

    \tikzstyle{box1} = [very thick, rectangle, rounded corners=1, minimum width=2.5cm, minimum height=2.2cm, text width=2.5cm, draw=darkgray!40, fill=gray!3, align=left]
    \tikzstyle{box2} = [very thick, rectangle, rounded corners=1, minimum width=2.5cm, minimum height=2.2cm, text width=2.5cm, draw=darkgray!40, fill=blue!5, align=left]
    \tikzstyle{arrow} = [very thick,->,>=stealth, darkgray]

    \begin{tikzpicture}

    \footnotesize
    \node(01) [box1] {\normalsize 1.\\Identify\\Policy Roster\\Actors};
    \node(02) [box1, right of = 01, xshift = 2.2cm] {\normalsize 2.\\Identify\\Organization\\Primary};
    \node(03) [box1, below of = 02, yshift = -2.0cm] {\normalsize 3.\\Identify\\Individual\\Executive};
    \node(04) [box2, right of = 02, xshift = 2.2cm] {\normalsize 4.\\Identify\\Side\\Accounts};
    \node(05) [box1, right of = 04, xshift = 2.2cm] {\normalsize 5.\\Classify\\Side\\Accounts};
    \node(06) [box2, below of = 04, yshift = -2.0cm] {\normalsize 6.\\Merge\\Account\\Data};
    \node(07) [box2, right of = 06, xshift = 2.2cm] {\normalsize 7.\\Collect\\Twitter\\Data};

    \scriptsize
    \draw [arrow] (01) -- (02);
    \draw [arrow] (01) -- (03);
    \draw [arrow] (02) -- (04);
    \draw [arrow] (04) -- (05);
    \draw [arrow] (02) -- (06);
    \draw [arrow] (03) -- (06);
    \draw [arrow] (05) -- (06);
    \draw [arrow] (06) -- (07);

    \end{tikzpicture}
    \caption{Data collection protocol. Gray tasks are fully manual and blue tasks are computer-assisted.} 
    \label{fig:protocol} 
\end{figure}

\paragraph{Organization Primary}
This is the primary organizational account of the policy actor. There is only one account in this category for each organization. This account can be manually identified in a number of ways, including from the organization's website, from a Twitter search, or from a general web search.

\paragraph{Individual Executive}\label{sec:indvmain}
This level includes the executive personnel of the organization. We consider these as individuals with decision-making power for the entire organization. This specifically means that we want to collect leaders of the top-level unit in the organization (i.e., before organizational subunits become parallel). Leaders of lower-level parallel subunits should not be included. For example, in a business organization, these accounts would be for members of the top level executive management team.

Individual executive accounts can be manually identified in a number of ways, including from the organization's website, from a Twitter or Linkedin search, or a general web search.

\paragraph{Organization Auxiliary}
This level includes the accounts of organizational subunits or accounts used for the organization's focused functions. Because we have a high number of policy actor organizations, we have developed a Twitter-specific protocol to collect this data. Specifically, we only include Twitter accounts that satisfy all of the following criteria:

\begin{enumerate}[noitemsep]
    \item Follows the organization's primary Twitter account
    \item Is followed by the organization's primary Twitter account
    \item Has at least one of the organization's pre-specified keywords in its Twitter bio; these organization-specific keywords are specified by country experts, and usually include at least different variations of the organization's name
    \item Twitter bio passes manual filtering
\end{enumerate}

Tasks 1--3 are computer assisted. For each organization's primary account, we used Twitter's API to collect the list of accounts it followed and the list of accounts that followed it. Then, we collected the Twitter bios of accounts satisfying both conditions to match against the organization's keywords. Finally, the remaining accounts are manually checked by a country expert.

\paragraph{Individual Non-executive}\label{sec:indivside}
This level includes all identifiable individuals that work or are members of the organization and have not already been identified as an organization executive. Again, because this is a difficult category to identify, we used the same criteria as the ones we used for identifying organization auxiliary accounts. In the manual filtering stage, accounts that have already been identified as an individual executive account are naturally excluded.

Depending on the type of organization, drawing the line between the executives and non-executives can be difficult. Some organizations have a relatively flat structure without a clear leader or executive team (e.g., social movement organizations like Extinction Rebellion). In such cases, we labelled all their representatives as executive personnel.

\subsection{Data Collection and Preprocessing}\label{sec:data}
After identifying the complete roster of accounts, we collected, for each account, all Twitter activity (or `statuses') for January 1, 2017--December 31, 2022. This included all original tweets with or without mentions of other accounts, all replies to other tweets, and all retweets. We did not collect data on `likes'. We collected the data using the \texttt{statuses/filter} endpoint from Twitter's V1.1 API suite, specifically matching by the users' ID or username. Data collection took place at several different time points during December 2021--February 2023.
Across the nine countries, we collected a total of 47.7 million statuses.

\subsubsection{Subsetting to Climate Change Statuses}\label{sec:dictionary}
Given our focus, we further subset our data to climate change-related tweets using a relative strict dictionary-based approach, summarized in \autoref{fig:climate_subset}. We describe these steps in detail in \autoref{app:climate_subset}.

\begin{figure}[!htb] 
    \centering 

    \tikzstyle{box1} = [very thick, rectangle, rounded corners=1, minimum width=3.2cm, minimum height=2.2cm, text width=3.2cm, draw=darkgray!40, fill=gray!3, align=left]
    \tikzstyle{arrow} = [very thick,->,>=stealth, darkgray]

    \begin{tikzpicture}

    \footnotesize
    \node(01) [box1] {\normalsize 1.\\Preprocessing\\Tweet Text};
    \node(02) [box1, right of = 01, xshift = 3.0cm] {\normalsize 2.\\Language-aware\\Lemmatization};
    \node(03) [box1, right of = 02, xshift = 3.0cm] {\normalsize 3.\\Climate-content\\Matching};
    \node(04) [box1, right of = 03, xshift = 3.0cm] {\normalsize 4.\\Match\\Validation};
    \scriptsize
    \draw [arrow] (01) -- (02);
    \draw [arrow] (02) -- (03);
    \draw [arrow] (03) -- (04);

    \end{tikzpicture}
    \caption{Climate-content labelling pipeline.} 
    \label{fig:climate_subset} 
\end{figure}

\clearpage
\singlespacing\footnotesize
\section*{CRediT Author Statement}
\textbf{Ted Hsuan Yun Chen:} Conceptualization, Methodology, Software, Validation, Formal analysis, Investigation, Data Curation, Writing - Original Draft, Visualization, Supervision.
\textbf{Arttu Malkamäki:} Conceptualization, Methodology, Software, Validation, Investigation, Data Curation, Writing - Original Draft, Supervision.
\textbf{Ali Faqeeh:} Conceptualization, Software, Validation, Formal analysis, Investigation, Writing - Original Draft.
\textbf{Esa Palosaari:} Conceptualization, Validation, Investigation, Data Curation.
\textbf{Anniina Kotkaniemi:} Conceptualization, Methodology, Validation, Investigation, Data Curation, Writing - Original Draft.
\textbf{Hasti Narimanzadeh:} Conceptualization, Methodology, Validation, Investigation, Writing - Original Draft.
\textbf{Laura Funke:} Data Curation.
\textbf{Cáit Gleeson:} Data Curation.
\textbf{James Goodman:} Validation, Resources, Data Curation.
\textbf{Antti Gronow:} Conceptualization, Validation, Data Curation, Writing - Review \& Editing, Resources.
\textbf{Marlene Kammerer:} Validation, Resources, Data Curation.
\textbf{Myanna Lahsen:} Validation, Resources, Data Curation.
\textbf{Alexandre Marques:} Data Curation.
\textbf{Petr Ocelik:} Validation, Data Curation, Resources.
\textbf{Shivangi Seth:} Validation, Data Curation.
\textbf{Mark Stoddart:} Validation, Resources, Data Curation.
\textbf{Martin Svozil:} Validation, Data Curation.
\textbf{Pradip Swarnakar:} Validation, Resources, Data Curation.
\textbf{Matthew Trull:} Data Curation.
\textbf{Paul Wagner:} Validation, Resources, Data Curation.
\textbf{Yixi Yang:} Validation, Data Curation.
\textbf{Mikko Kivelä:} Conceptualization, Validation, Resources, Writing - Review \& Editing, Supervision, Project administration, Funding acquisition.
\textbf{Tuomas Ylä-Anttila:} Conceptualization, Validation, Resources, Data Curation, Writing - Review \& Editing, Supervision, Project administration, Funding acquisition.

\bibliography{policytwitter}

\clearpage
\normalsize

\appendix

\newcounter{sisection}
\setcounter{sisection}{1}

\renewcommand{\thesection}{S{\arabic{sisection}}}
\renewcommand{\thefigure}{\thesection.\arabic{figure}}
\renewcommand{\thetable}{\thesection.\arabic{table}}


\counterwithin{figure}{section}
\counterwithin{table}{section}

\clearpage

\section{Subsetting to Climate Tweets}\label{app:climate_subset}
To identify which of our tweets are related to climate change, we applied a multi-stage dictionary approach to all collected Twitter statuses, summarized in \autoref{fig:climate_subset_sup}. Following the last stage, all statuses that had at least one word matched to our dictionary are marked as a climate-relevant tweet.

\begin{figure}[!b] 
    \centering 

    \tikzstyle{box1} = [very thick, rectangle, rounded corners=1, minimum width=3.2cm, minimum height=2.2cm, text width=3.2cm, draw=darkgray!40, fill=gray!3, align=left]
    \tikzstyle{arrow} = [very thick,->,>=stealth, darkgray]

    \begin{tikzpicture}

    \footnotesize
    \node(01) [box1] {\normalsize 1.\\Preprocessing\\Tweet Text};
    \node(02) [box1, right of = 01, xshift = 3.0cm] {\normalsize 2.\\Language-aware\\Lemmatization};
    \node(03) [box1, right of = 02, xshift = 3.0cm] {\normalsize 3.\\Climate-content\\Matching};
    \node(04) [box1, right of = 03, xshift = 3.0cm] {\normalsize 4.\\Match\\Validation};
    \scriptsize
    \draw [arrow] (01) -- (02);
    \draw [arrow] (02) -- (03);
    \draw [arrow] (03) -- (04);

    \end{tikzpicture}
    \caption{Climate-content labelling pipeline.} 
    \label{fig:climate_subset_sup} 
\end{figure}

\paragraph{Preprocessing}
In the first stage, we started by conducting the following text preprocessing procedures to each status:
\begin{enumerate}[noitemsep]
    \item Concatenated tweets and referenced tweets: As responses to climate content is likely to be climate related as well, we included referenced tweets (i.e., quoted tweets) into the status's text string
    \item Removed all URLs: This helps remove false positive matches for short keywords such as `CO2'
    \item Split camel cases: As it is common to write hashtags and mentions in camel case, we split camel case occurrences into separate strings then inserted them into the status's text string
    \item Removed all punctuation and symbols
\end{enumerate}

\paragraph{Lemmatization} In the second stage, we lemmatized all our preprocessed statuses.
As some of the keywords can take different forms (e.g., inflections) depending in the language and usage, lemmatization is needed to make sure these different forms can be captured when matching to the selected keywords. For example, in Finnish, lemmatization is required for several inflections used to refer to global warming, e.g., `maapallolla lämpenemisellä', to match our keyword `maapallo lämpeneminen'.

Each status was lemmatized once for English and once (where relevant for specific countries) for the language most spoken or used by policy actors in that national policy system. Specifically, the following countries had additional non-English lemmatization: Czechia (Czech), Finland (Finnish), Canada (French), Germany (German), India (Hindi), Brazil (Portuguese), and Sweden (Swedish).
We used the following language-appropriate libraries for this procedure. We used \texttt{spaCy} because of its high reported accuracy measures for all languages except for Czech (which was absent from the library) and Hindi (for which lemmatization was not reliable). For Czech, we used \texttt{Simplemma} (\url{https://pypi.org/project/simplemma}), which we investigated and verified its reliable performance on a sample of tweets. For Hindi, we used the pre-processed text without performing lemmatization.

\begin{table}[h!]
\centering\tiny
\renewcommand{\arraystretch}{1.5}
\begin{tabularx}{\textwidth}{X X X X X X X}
\hline\hline
\multicolumn{3}{l}{\textbf{Common words}} \\ \hline
cop23 & cop24 & cop25 & cop26 & cop27 & ipcc  & unfccc \\ co2  \\ \\ \hline

\textbf{English} & \textbf{Finnish} & \textbf{Swedish} & \textbf{German} & \textbf{Czech} & \textbf{Portuguese} & \textbf{French} \\ \hline
\rowcolor{mylightgray} climate & ilmasto & klimat & klima & klima & clima/climátic & climat \\
emission & päästö & utsläpp & emission & emise & emissão & émission \\
\rowcolor{mylightgray} fossil & fossiili & fossil & fossil & fosilní & fóssil & fossile \\
carbon & hiili & kol & kohlenstoff & uhlík & carbono & carbone \\
\rowcolor{mylightgray} coal & hiili & kol & kohle & uhlí & carvão & charbon \\
greenhouse & kasvihuone & växthus & treibhaus & skleníkový & estufa & serre \\
\rowcolor{mylightgray} global\linebreak warming  & maapallon\linebreak lämpeneminen & global \linebreak uppvärmning & globale\linebreak erwärmung & globální oteplování & aquecimento global & réchauffement climatique \\
\rowcolor{mylightgray}  & maapallosta\linebreak lämpeneminen  & & & globální oteplený &  & réchauffemer\linebreak climatique  \\ \hline
\end{tabularx}
\caption{Keywords used for each language, excluding Hindi keywords which are included in \autoref{tab:keywords_hindi}.}
\label{tab:keywords}
\end{table}

\paragraph{Matching} In the third stage, we matched our lemmatized statuses with the climate keyword dictionary in \autoref{tab:keywords} following these specifications: 1) each status is checked for keyword matches in English and its additional language where relevant, 2) only the start of the word string needs to match, i.e., wildcards are added to the end of the keywords, 3) case insensitive, and 4) aware of accents on letters. For each status, all words that matched to any keyword are extracted as the (keyword, matching word) pair for validation.

\begin{table}[h!]
\centering\footnotesize\renewcommand{\arraystretch}{1.2}

\caption{Classification Performance by Country}
\label{tab:performance-by-country}
\begin{tabular}{l c c c c c}
\hline\hline
{Country}        & {Accuracy} & {Recall} & {Specificity} & {Precision} & {F1-Score} \\ \hline
Australia               & 0.78              & 0.84            & 0.56                 & 0.90               & 0.87              \\
Ireland                 & 0.97              & 0.99            & 0.83                 & 0.98               & 0.98              \\
Finland                 & 0.94              & 0.97            & 0.50                 & 0.97               & 0.97              \\
Sweden                  & 0.95              & 0.97            & 0.86                 & 0.96               & 0.97              \\
Germany                 & 0.86              & 0.96            & 0.09                 & 0.89               & 0.92              \\
Canada                  & 0.92              & 0.92            & 0.94                 & 0.99               & 0.95              \\
Brazil                  & 0.93              & 0.94            & 0.89                 & 0.97               & 0.96              \\
Czech Republic          & 0.91              & 1.00            & 0.62                 & 0.91               & 0.95              \\
India                   & 0.93              & 0.97            & 0.50                 & 0.96               & 0.96              \\ \hline
\end{tabular}
\end{table}

\paragraph{Validation}
In the last stage, the (keyword-matched word) pairs obtained from the matching stage are validated using a combination of manual human inspection and LLM classification. First, we sorted the pairs in descending order by the number of statuses that they appear in. Starting from the first pair, we calculated the cumulative proportion of all matched statuses are captured by the current and all preceding keywords, as well as the current keywords' marginal contribution to the cumulative proportion.

We then set a threshold of pairs to manually examine based on the cumulative matched proportion (and after removing all pairs that had no marginal contribution). For six countries, we set the threshold to $0.99$, but $0.95$ for Finland, Germany, and Sweden due to their massive set of pairs stemming from the frequency of compound words in these languages. The remaining pairs that were not inspected by human reviewers are then passed to a large language model for classification. Based on validation with our manually inspected test set, we chose the Llama-3.3-70B-Instruct model (see \autoref{tab:performance-by-country} for its classification performance). Our prompt and implementation is outlined in \autoref{fig:llm_prompt}.

\begin{figure}[!h]
\centering
\begin{lstlisting}[breaklines,basicstyle=\scriptsize\ttfamily]
System prompt:
==============
You are a Twitter user who has seen contents with different context, for
example, casual, economical, political, or societal topics but are only
interested in posts that concern climate change and not posts in other
contexts.

User prompt:
============
You are familiar with online discussions about causes and effects of climate
change in Twitter and the related popular topics, campaigns, memes, and
hashtags. Consider the following terms in [language], and for each determine
that in the case the term was used on a post in Twitter, if it is very likely
that the post was related to a climate change topic (label 1) or likely to be
any other topic (label 0). If unclear, choose 2.

[terms block]

Please do not generate any other text and make **only** an output that is in
a valid JSON format identical to that of the following examples:

[
 { "term":"coalition",        "decision":0, "explanation":"..."},
 { "term":"carbon",           "decision":1, "explanation":"..."},
 { "term":"carbonara",        "decision":0, "explanation":"..."},
 { "term":"climatechnage",    "decision":1, "explanation":"..."},
 { "term":"coal",             "decision":1, "explanation":"..."},
 { "term":"climateemergence", "decision":1, "explanation":"..."},
 [other examples]
]

Such that:
   - term: the given term
   - decision: 1 if very much climate-related rather than another context,
     0 if rather likely to be another context, or 2 if unclear
   - explanation: one or a few sentences justifying the decision

Remember to check that the output has the above correct JSON format and the
end of the file is closed properly. That is, output exactly [number of terms]
JSON objects (one per term) and make sure to close the array according to
correct JSON format.

Now classify each of the given terms.

\end{lstlisting}\caption{LLM prompt. In the prompt, \texttt{[terms block]} is a subset of the terms passed to the model in each round, \texttt{[other examples]} included examples sampled from the words tagged by humans, \texttt{[number of terms]} was the size of terms subset, and \texttt{[language]} was the second language for the corresponding country.
The model usually needed several generation attempts to return the results correctly according to prompted requirements. Missed terms were resupplied to the model until tags were generated or at least 50 attempts were made. In the rare cases exceeding 50 attempts, the terms were dropped from tagging.}\label{fig:llm_prompt}
\end{figure}

\clearpage
\begin{figure}[h!]
    \centering
    \includegraphics[width=0.45\linewidth]{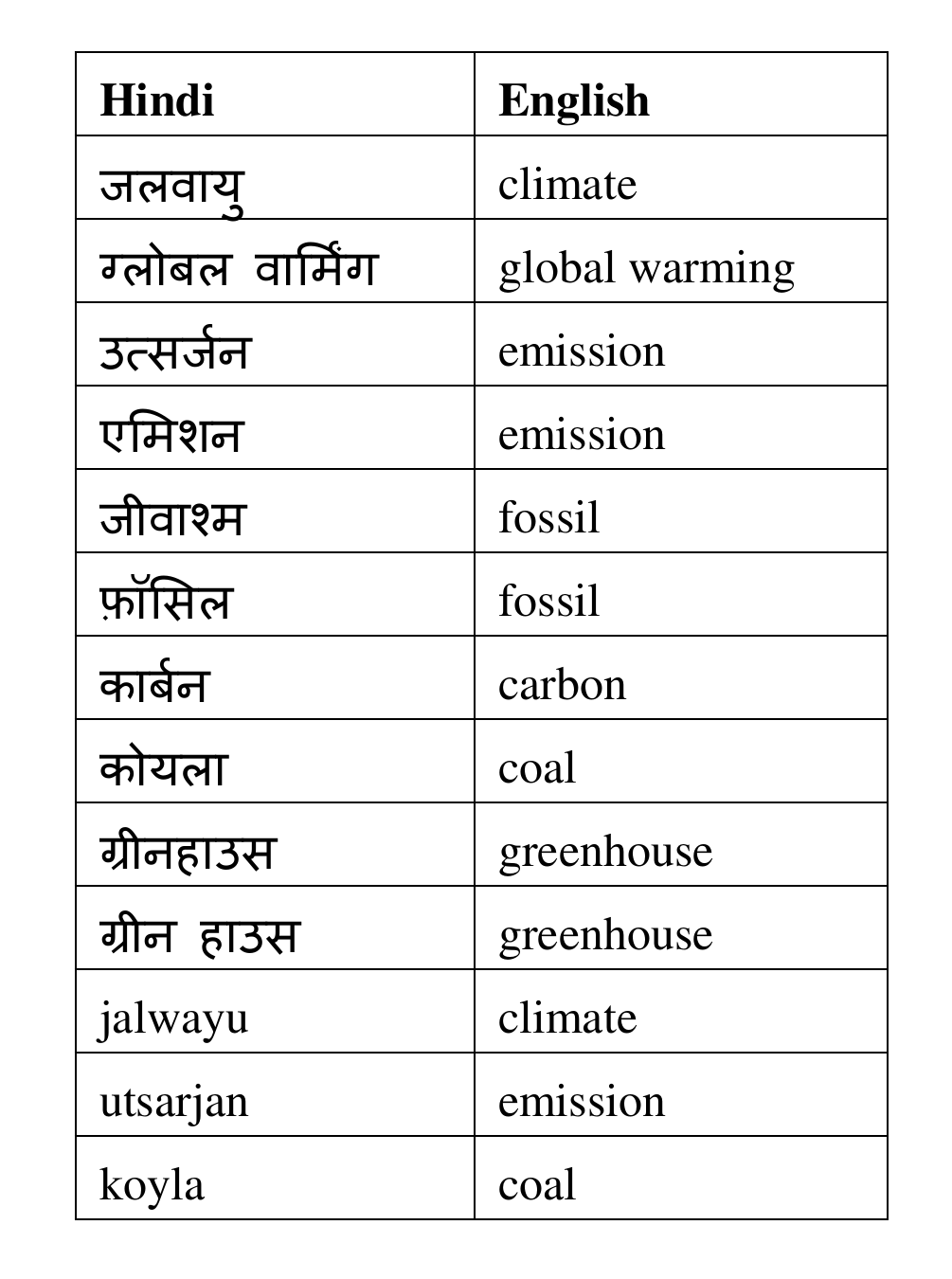}
    \caption{Hindi Keywords and their translations.}
    \label{tab:keywords_hindi}
\end{figure}

\clearpage
\setcounter{sisection}{2}
\section{Position Inference}\label{app:position_inference}
We divided the policy actors into groups with opposite climate change policy positions with a two-step protocol. First, we applied a graded response model to survey answers collected from policy actor organizations in seven of our nine case countries: Australia, the Czech Republic, Finland, Germany, India, Ireland and Sweden. This analysis yielded values indicating the climate policy positions for organizations that had responded to the surveys. Second, the values were then used in a position inference task for organizations that are a part of the policy networks but had not responded to the survey, including for all organizations in countries where we did not conduct surveys, i.e., Brazil and Canada. We inferred these additional climate policy positions using the network lasso approach, which propagated the survey-derived positions to other policy actors based on how close they are in our collected retweet network.

\subsection{Measuring Climate Policy Positions with Survey Data}

In seven of the nine countries, we surveyed the policy actor organizations about their climate policy positions. The survey was sent to the persons responsible for climate or environmental policy in the organizations, or in the rare absence of such personnel, to the person with an overview of the whole organization. To maximize the response rate, we contacted the respondents by phone in advance, and sent several reminders to encourage participation. In the survey, the respondents were asked to provide answers that best represent their organizations' views. The questions concerned, for example, the veracity of climate change, trust in climate science, and preferences on climate change policies. The countries differ in the questions they included in their surveys, but common items are present in all countries. \autoref{tab:surveyinfo} shows the years of our survey data collection.

\begin{table}[h!]
\centering\footnotesize\renewcommand{\arraystretch}{1.2}
\caption{Survey information}
\label{tab:surveyinfo}
\begin{tabular}{l c}
\hline\hline
{Country}        & {Survey Year} \\ \hline
Australia               & 2024  \\
Czech Republic          & 2025 \\
Finland                 & 2020 \\
Germany                 & 2022 \\
India                   & 2022 \\
Ireland                 & 2021 \\
Sweden                  & 2022 \\ \hline
\end{tabular}
\end{table}

\begin{figure}
    \centering
    \includegraphics[width=0.5\linewidth]{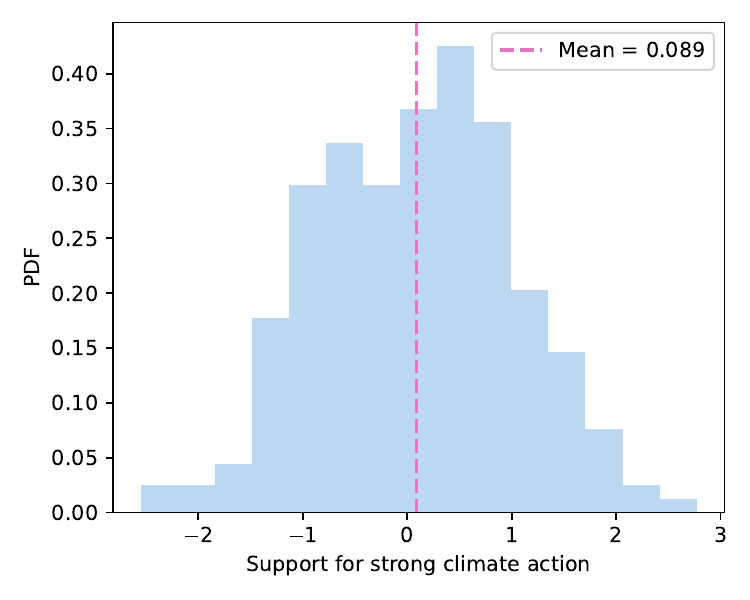}
    \caption{Distribution of the surveyed climate change policy positions across 7 countries. There are 444 surveyed positions.}
    \label{fig:inferred_positions}
\end{figure}

We analyzed the organizations' climate change policy positions expressed in the survey using a graded response model \citep{samejima2016graded}. Graded response models belong to the family of item response models that apply a probabilistic function to Likert-scale answers, given a `latent trait' that is assumed to affect the probability of the respondent choosing a specific answer category. It is denoted with $\theta$, a continuous value centered around a mean of zero, ranging from negative to positive. Respondents with higher levels of $\theta$ are more likely to endorse answer categories representing a higher intensity of the assumed latent trait, and vice versa.  The latent trait we expect to affect how respondents answer questions about climate change policy preferences is their overall support for strong climate action. Thus, positive values indicate climate action support, whereas negative values show the opposite. In total, we inferred the climate policy positions of 444 organizations. Because these surveys are designed to capture the organization's position instead of the position of the individual respondent, we apply the inferred positions to the given organization's primary Twitter account. \autoref{fig:inferred_positions} shows the distribution of our inferred positions.

\subsection{Propagating Climate Policy Positions through Networks Ties}\label{sec:lasso}

Using the climate policy positions inferred from our surveys, we further inferred, via a label propagation process, the positions of all remaining Twitter accounts, i.e., the primary accounts of organizations that did not respond to the survey and accounts at all other organization levels.

A common approach to inferring labels in network data is based on the notion that nodes in a similarity or affinity network tend to exhibit similar characteristics. Labels observed for a subset of nodes can propagate through the network to unlabeled nodes. The network lasso formalizes this idea as an optimization problem that penalizes differences between connected nodes, thereby encouraging similar estimates for nodes linked by strong edges, while allowing for heterogeneity across weakly connected regions~\citep{tibshirani2005sparsity, hallac2015network}.

The network lasso problem is conventionally defined \citep{hallac2015network} as follows:
\begin{equation}\label{eq:hallac-lasso}
\min_{x}
\sum_{i\in V}f_i(x_i)
+
\lambda \sum_{(i,j)\in E} w_{ij}\,\lvert x_i - x_j \rvert ,
\end{equation}
where $E$ is set of edges of the network, $w_{ij}$ is weight of the directed edge from node $i$ to $j$, $x_i$ the latent scalar value of node $i$, and $f_i(.)$ a cost function at node $i$.

In the standard network lasso formulation, the data-fit term $f_i(x_i)$ softly anchors the estimations at labeled nodes to their observed labels, while the graph regularization term, i.e.~edge penalty, promotes similarity between adjacent nodes and propagates label information to unlabeled nodes. This soft anchoring allows the estimated values at labeled nodes to deviate from their observations when such deviations lead to a more coherent graph-regularized solution~\citep{zhou2003learning}.

This formulation is especially useful in settings where the observed labels may be noisy, incomplete, or where the network is intended to correct local inconsistencies by enforcing agreement with adjacent nodes. Here, the labeled values are scores based on survey responses, which we treat as the best available measurements of the responding organizations' climate policy positions. Allowing the optimization to alter them would mean that the structure of the network could override these observed survey-based positions. We therefore hold the labeled nodes fixed and use them as boundary conditions. In doing so, the graph regularization term propagates the position information to unlabeled nodes without perturbing the survey-based estimates.\footnote{Additionally in the analyses we conducted, the conventional formulation with soft constraints performed significantly worse at predicting the position of held-out nodes in the hold-one-out experiments.}

Thus, we estimated latent climate change policy position values via solving the following convex, non-smooth variant of the network lasso with hard constraints on the observed nodes:
\begin{equation}\label{eq:lasso}
\min_{x}
\sum_{(i,j)\in E} w_{ij}\,\lvert x_i - x_j \rvert
\quad \text{subject to} \quad x_s = y_s \;\; \forall s \in S,
\end{equation}
where $S$ denotes the set of seed nodes, i.e.,~the nodes for which climate positions are available from the survey responses, and $y_s$ is the corresponding observed position score for seed node $s$. Because the seed-node values are imposed as hard constraints rather than through a data-fit penalty, there is no regularization parameter controlling a trade-off between label agreement and graph smoothness. The optimization fixes the seed nodes at their exact observed positions, and estimates the positions of the unlabeled by minimizing the weighted variation over the graph. Any positive scalar multiplier on the objective would leave the minimizer unchanged.

We solved the network lasso via an edge-splitting ADMM (Alternating Direction Method of Multipliers) scheme, introducing local copies of endpoint values per each directed edge, following the augmented Lagrangian and update rules outlined in~\citep{hallac2015network}, adapted to our constraints in Eq.~\eqref{eq:lasso}. Each iteration of our implementation of the ADMM algorithm updates latent scalar node values $x_i$ and two intermediary variables in tandem, while ensuring the latent values of the seed nodes are always clamped to their observed values.

Label propagation generally requires a similarity or affinity network. Here, we used the retweet network, as retweets are indicators of agreement or endorsement \citep{metaxas2015retweets}.
As a preprocessing step, we restricted the network to the accounts included in the nine-country roster and kept only the induced directed network on those nodes. The resulting largest weakly connected component retained 424 seed nodes. Edge weights are raw retweet counts, that is, there is an edge from node $i$ to $j$ with weight $r$ if node $i$ has retweeted $j$ $r$ times. Edges with weights 0 were discarded. The number of seed nodes per country remaining after the preprocessing steps is reported in \autoref{tab:seed-info}.

\begin{table}[h!]
\centering\footnotesize\renewcommand{\arraystretch}{1.2}

\caption{Seed node counts and fractions by country}
\label{tab:seed-info}
\begin{tabular}{l c c c}
\hline\hline
{Country} & {Nodes} & {Seed nodes} & {Seeded (labeled) \%} \\ \hline
Australia & 633 & 42 & 0.066 \\
Brazil & 574 & 0 & 0.000 \\
Canada & 2530 & 0 & 0.000 \\
Czech Republic & 353 & 47 & 0.133 \\
Germany & 2138 & 39 & 0.018 \\
Finland & 4534 & 85 & 0.019 \\
Ireland & 804 & 72 & 0.090 \\
India & 1015 & 65 & 0.064 \\
Sweden & 1811 & 74 & 0.041 \\ \hline
\end{tabular}
\end{table}

\begin{figure}[h!]
    \centering
    \includegraphics[width=\linewidth]{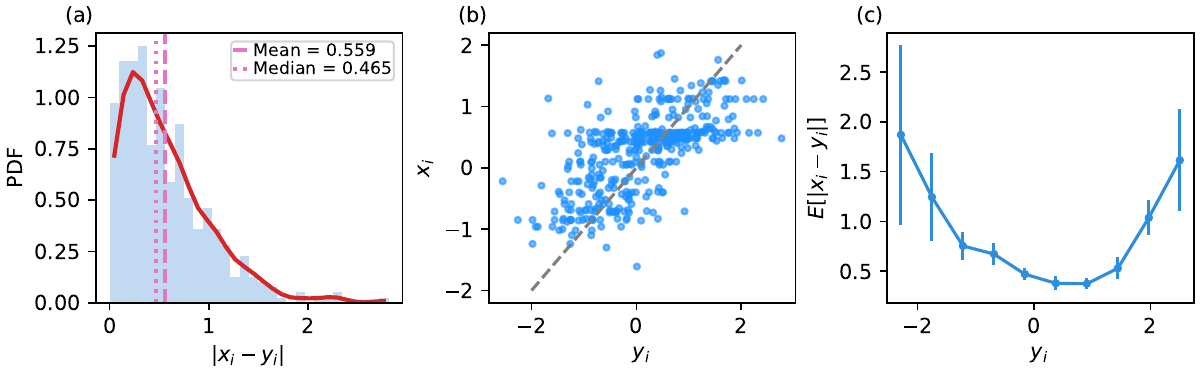}
    \caption{Hold-one-out validation of the hard-constrained network lasso on seed nodes. Panel (a) shows the probability density function of the absolute prediction errors; panel (b) compares the inferred and observed seed values; and panel (c) displays the binned mean absolute error as a function of the observed seed value. Error bars show 95\% confidence intervals.}
    \label{fig:hoo}
\end{figure}

To assess the predictive performance we conducted hold-one-out validation over the seed nodes. In each validation run, one seed node was held out and the scalar positions for the rest of the network were estimated via minimizing the objective function in Eq.~\eqref{eq:lasso} using the remaining seed nodes. The validation results comparing held-out estimations to observed seed values are illustrated in \autoref{fig:hoo}. As reported in \autoref{fig:hoo}(a), the held-out node had on average an absolute difference of 0.559 between its inferred and survey positions, with \autoref{fig:hoo}(c) indicating that nodes at the two extremes of the climate policy position scale have on average higher values of absolute errors.

\begin{figure}[h!]
    \centering
    \includegraphics[width=0.9\linewidth]{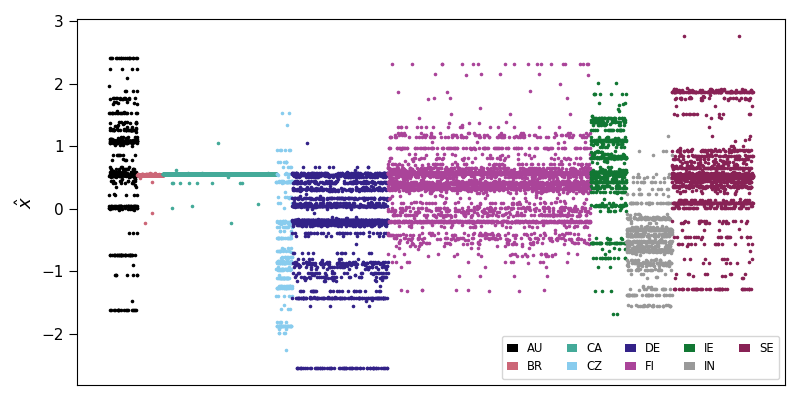}
    \caption{Distribution of inferred climate-policy positions of accounts across all nine countries. The two countries with no seed values (Brazil and Canada) show the lowers variation among nodes. The same data in network visualization form can be seen in \autoref{fig:lasso-network-vis}.}
    \label{fig:inferences-dist}
\end{figure}

\begin{figure}[h!]
    \centering
    \includegraphics[width=0.85\linewidth]{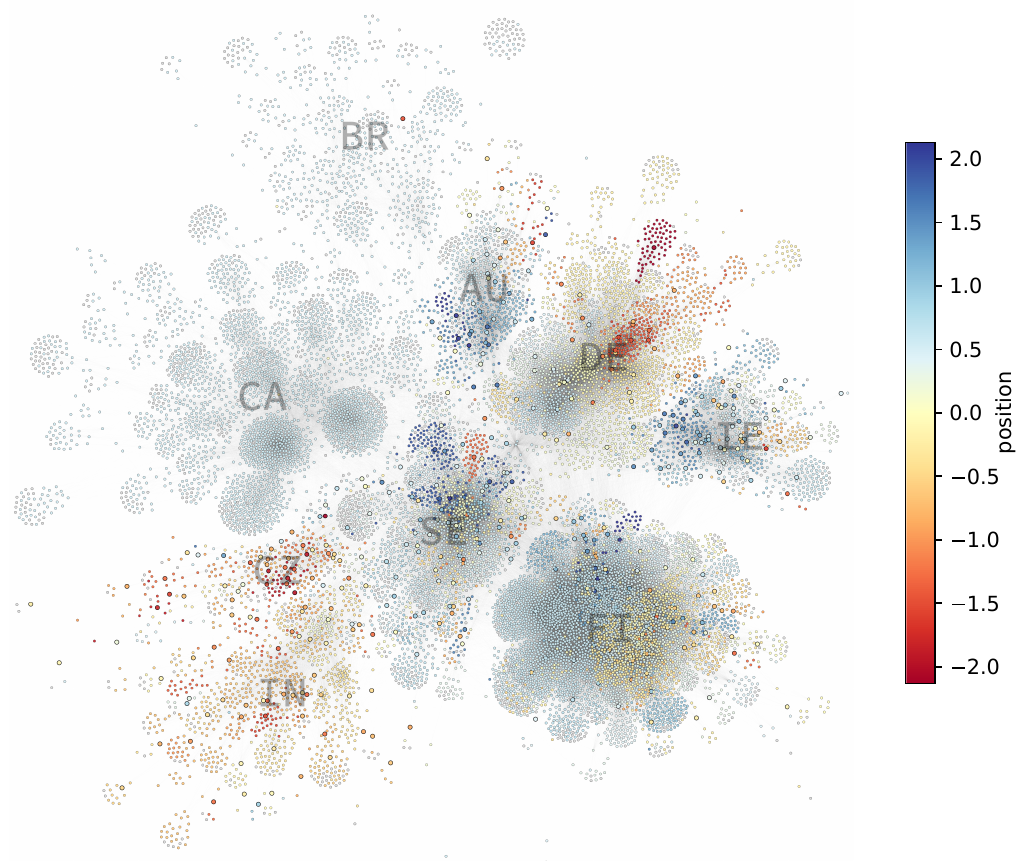}
    \caption{Retweet network of all interactions, climate or otherwise, 2017-2022. Node color indicates inferred climate-policy position. Only the largest weakly connected component is shown. The seed nodes are shown as larger nodes with heavier borders. The color representation of position values are clipped to the range -2.13 to 2.13 interval (which contains more than 95\% of nodes) to show the inter-node variability while also preserving color map symmetry. The distribution of inferred values (before clipping) can also be seen in \autoref{fig:inferences-dist}.}
    \label{fig:lasso-network-vis}
\end{figure}

Additionally, \autoref{fig:inferences-dist} shows the inferred position of all accounts by country. For the two countries with no survey results, the estimation procedure reduces to using the average best estimate for all but a few nodes. In other countries, however, the estimates form clusters of nodes with close values, due to the nature of the $L_1$ loss function in Eq.~\eqref{eq:lasso}. \autoref{fig:lasso-network-vis} displays these estimated positions on the retweet network, showing that clusters of highly connected nodes tend to have similar estimated positions.

To evaluate the stability of the inferred positions with respect to variation in retweet counts, we performed edge-weight bootstraps. In each bootstrap replicate, the node set and observed edge support were kept fixed, while retweet counts were resampled using a multinomial draw. We conducted two variants that differ by the constraint placed on the resampling probability. In a per-node variant, each node's total outgoing retweet count was redistributed across its observed outgoing edges in proportion to the original edge weights. In a global variant, all retweet counts were redistributed proportional to the original weights. In each case, we then re-estimated the network lasso positions on each bootstrapped network using the same hard seed constraints.

The per-node and global bootstraps capture slightly different sources of uncertainty in the retweet weights. The per-node variant preserves each account’s total outgoing retweet volume and varies only how that volume is distributed across its observed outgoing edges. The global variant relaxes this constraint, allowing retweet weight to be redistributed across the network as a whole. The resulting intervals provide a robustness check for the observed network interactions rather than uncertainty in the survey-derived seed positions.

\begin{figure}[h!]
    \centering
    \includegraphics[width=0.9\linewidth]{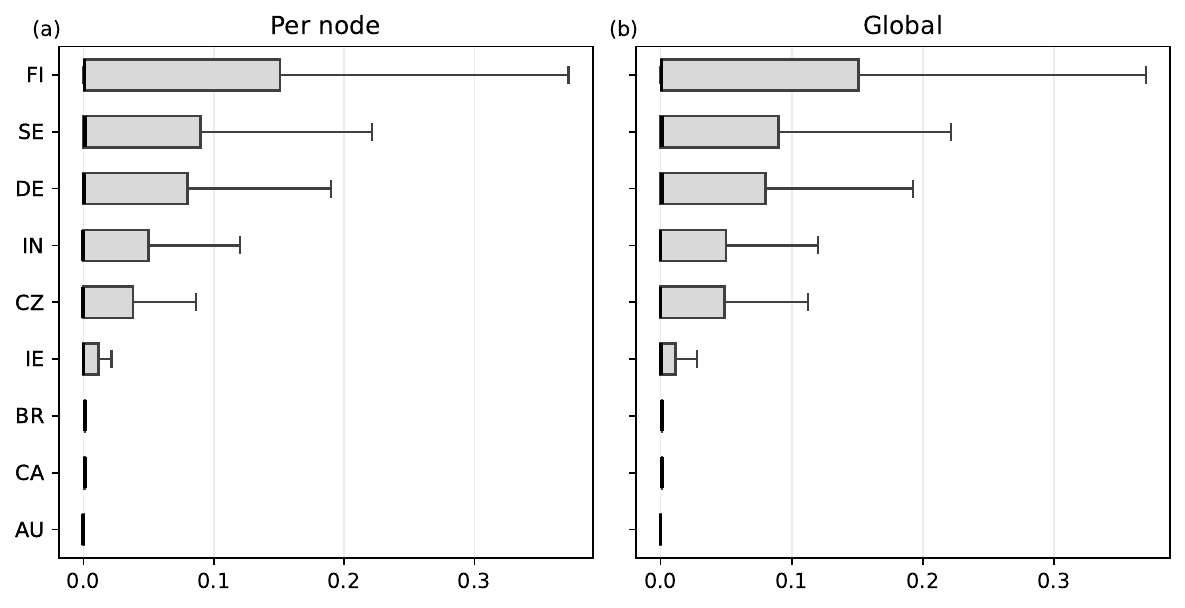}
    \caption{Bootstrap stability of network-lasso position estimates by country. For each bootstrap replicate, retweet edge weights were resampled and network-lasso positions were re-estimated. The boxplots show the distribution, across accounts in each country, of node-level 95\% bootstrap interval widths. (a) The per-node variant, where each node's outgoing retweet count is redistributed across its observed outgoing edges. (b) The global variant, where retweet counts are redistributed across all observed edges. Larger interval widths indicate greater sensitivity of inferred positions to perturbations in retweet weights.}
    \label{fig:lasso-bootstrap}
\end{figure}

\autoref{fig:lasso-bootstrap} presents the edge-bootstrap results by country. For each node, we computed the width of its 95\% bootstrap interval, defined as the difference between the 97.5th and 2.5th percentiles of its bootstrap estimates. The figure then summarizes the distribution of the interval widths of all nodes in each country. Wider interval for a node indicates that its inferred position depends more strongly on the precise retweet counts. The results show that in every country the median 95\% interval width is very small, less than 0.0012 in all cases, indicating very low sensitivity of the estimates to variations in retweet counts. In the case of countries with no survey data (Brazil and Canada) the bootstrap interval widths are markedly narrow. This is due to the fact that the main uncertainty for these countries instead concerns the absence of within-country survey anchors, which is not captured by this bootstrap procedure.

\end{document}